\documentclass[preprint,showpacs,preprintnumbers,amsmath,amssymb]{revtex4}
\usepackage{epsfig,amsmath,amssymb,graphics,color,calc,wasysym}
\usepackage{ifpdf}

\begin{document}

\title[]{Measuring Colloidal Forces from Particle Position Deviations inside an Optical Trap}

\author{Djamel El Masri\footnote{J.H.M.alMasri@uu.nl}, Peter van Oostrum, Frank Smallenburg, Teun Vissers, \\Arnout Imhof, Marjolein Dijkstra, and Alfons van Blaaderen}

\affiliation{Soft Condensed Matter, Debye Institute for Nanomaterials Science, Princetonplein 5, 3584 CC Utrecht, The Netherlands.}

\date{\today}

\begin{abstract}
We measure interaction forces between pairs of charged PMMA colloidal particles suspended in a relatively low-polar medium (5 $\lesssim \varepsilon \lesssim$ 8) directly from the deviations of particle positions inside two time-shared optical traps. The particles are confined to optical point traps; one is held in a stationary trap and the other particle is brought closer in small steps while tracking the particle positions using confocal microscopy. From the observed particle positions inside the traps we calculate the interparticle forces using an ensemble-averaged particle displacement-force relationship. The force measurements are confirmed by independent measurements of the different parameters using electrophoresis and a scaling law for the liquid-solid phase transition. When increasing the salt concentration by exposing the sample to UV light, the force measurements agree well with the classical DLVO theory assuming a constant surface potential. On the other hand, when adding tetrabutylammonium chloride (TBAC) to vary the salt concentration, surface charge regulation seems to play an important role.
\end{abstract}

\maketitle

\section{Introduction}

Understanding the interactions between (charged) particles is of great importance from a fundamental point of view to many natural, biological and industrial processes, which require control over the structure, stability and many other properties of a dispersion.\cite{Glotzer, Russel} In addition, due to their mesoscopic size, colloids provide an ideal experimental system for the investigation of questions related to many particle statistical mechanics in- and out-of-equilibrium and related to both structural and dynamical properties of condensed matter. It is experimentally possible to visualize and follow their positions in real time using microscopy techniques. The particles can then be tracked to reconstruct their individual trajectories. From these, the interaction forces can be calculated. According to the classical Derjaguin-Landau-Verwey-Overbeek (DLVO) theory, the interaction between stable pairs of charged colloidal particles suspended in a dielectric medium is a sum of a repulsive electrostatic part and a generally shorter ranged attractive Van der Waals part.\cite{Derjaguin-Landau, Verwey-Overbeek} However, when DLVO theory is quantitatively confronted with experiments, some refinements are required.\cite{Hsu-Langmuir1999, Behrens-Langmuir2000, Jonathan02, Ohki1999} For example, the presence of adsorption of charged species at the colloidal surface and the complex charging mechanisms in nonaqueous media with low polarity,\cite{Lyklema1968, Morrison-CollSurf1993} make it unclear as to what extent the electrostatic interactions can be described with a volume fraction and phase independent sum of pair potentials.\cite{Reinke, RoyallJCP2006} Measuring interparticle forces directly in these systems should help our understanding of the underlying physics. 

Different experimental techniques have been advanced to directly or indirectly measure forces acting between charged objects from a few femto-Newtons to sub-pico-Newtons. Total Internal Reflection Microscopy (TIRM) can measure the interaction forces between a microscopic sphere and a flat surface.\cite{Prieve} The Magnetic Chaining Technique (MCT) has been used to directly probe the force-distance profile between magnetic colloidal particles.\cite{Calderon,Li} Optical Tweezers (OT) provide a powerful tool to optically manipulate colloidal systems and have been widely used in recent years to investigate colloidal interactions.\cite{Sainis,Kegler,RobertsJCP07,Mittal,Polin} It is also possible to measure interaction forces by inverting the pair correlation function $g(r)$ in the case of weakly interacting systems using the Boltzmann distribution, $U(r)/k_{\mathrm B} T \equiv -\ln~[g(r)]$.\cite{Weiss,Behrens,RoyallJCP2006} 

In the present paper, we use optical tweezers and a Nipkow scanning disk confocal microscope to measure interaction forces between pairs of charged PMMA colloidal particles suspended in a relatively low polar medium (cyclohexyl chloride (CHC), $\varepsilon = 7.6$) at different ionic strengths, going from very low (purified solvent) to high ionic strength (added salt). Two particles are trapped, one is held in a stationary trap and the other particle is brought closer in small fixed steps. From the observed deviations of the particle positions inside the traps we calculate the interparticle forces.

\section{Experimental}

In our study, we used sterically stabilized and fluorescently labelled poly-methyl-methacrylate particles (PMMA)\cite{Royall-CondMatt} with diameter of $\sigma = 1.4$ $\mu$m and size polydispersity of 3$\%$. The refractive index of CHC is $n=1.462$ at $\lambda=1064$ nm, below that of the particles $n_{\mathrm p}=1.494$, which allows optical tweezer experiments. In this low-polar solvent, charge dissociation still occurs spontaneously,\cite{van der Hoeven} contrary to truly apolar media that require charge stabilizing surfactants.\cite{Hsu} Electrophoresis measurements in the dilute dispersion showed that the particles carried a positive charge of about $+500e$ (with $e$ the elementary charge). We studied systems at a volume fraction $\varphi < 0.001$, determined by particle tracking. Four different samples were explored. The first one consisted of a PMMA dispersion in purified CHC using a method described elsewhere.\cite{Pangborn} In fact, the method offers a very convenient way to quickly reduce the conductivity (from $\sim$ 1000 pS/cm to less than 20 pS/cm) and at the same time, alumina is often used as a desiccant to remove traces of water. Additionally, the purified solvent is stored with added molecular sieves (4 \AA ngstr\"{o}m, Acros Organics) that serve as an adsorbent. In the three other systems, the ion concentrations were increased by exposing the initial purified sample under an UV lamp at different exposure times. It is known that UV exposure facilitates the partial degradation of cyclohexyl halides into ions,\cite{LeunissenThesis} but it is difficult to quantify the salt concentration by conductivity measurements without knowing exactly which ions are generated. However, our measurements give an idea of how the different parameters in the same sample vary when exposed to UV light. The dried particles were initially mixed with the purified CHC and allowed to equilibrate for 48 hours before measurements were performed. The sample was then gently remixed, placed in a glass capillary of $0.1 \times 2.0$ mm inner dimensions (VitroCom) and sealed with Norland 68 UV glue. While curing the glue, the sample was covered with aluminium foil to prevent solvent degradation.

The measurement of particle interactions was performed using optical tweezers.\cite{Ashkin,Crocker96} We use a pair of optical tweezers formed by focusing a 1064 nm laser, using the same lens as used for imaging. We applied time-sharing using accousto-optic deflectors to generate two traps and to vary their positions.\cite{Vossen} Particle imaging was performed with a Nipkow-disk scanning confocal microscope (CSU10, Yokogawa) and recorded on a digital video camera (Evolution$^{\texttt{TM}}$ QEi) as in Ref.\cite{PeterLu2007}. A $100\times$ 1.4 NA oil immersion objective (Leica PLAN APO) was used. The particles were dyed with Rhodamine and were excited with a Millennia V diode-pumped laser beam ($\lambda_0$ = 532 nm). The trapped particles were located in the plane at about 14 $\mu$m above the bottom of the capillary glass wall to avoid possible effects of the sample boundaries. One of the traps was brought closer to the other one in small steps of about 500 nm every 2-3 seconds. For every step, 1000 images of $80 \times 21$ pixels were recorded to sample the Brownian motion of the particles inside the traps. All images were processed to extract particle positions using home-made software based on methods similar to that described by Crocker and Grier.\cite{Crocker96}

\begin{figure}
\includegraphics[width=0.6\linewidth]{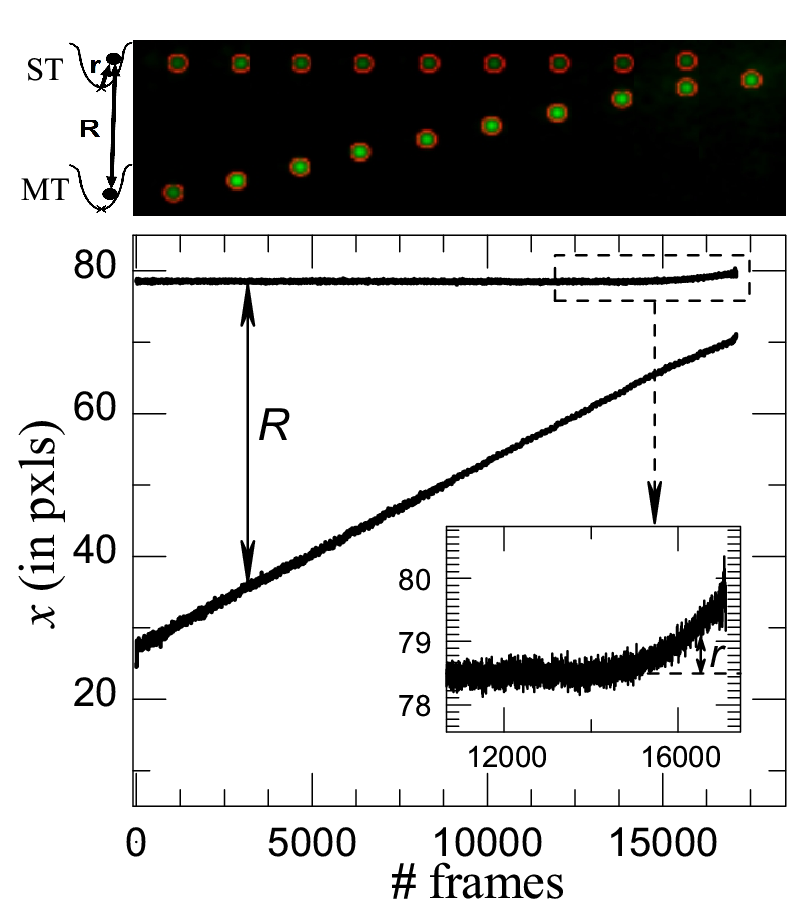}
\caption{(Top left) A sketch of two particles confined to two optical traps. (Top right) Confocal microscopy images of the two particles as the mobile trap (MT) approaches the stationary one (ST). The red contours on the particles are centered on the positions taken from the tracking code. (Bottom) Trajectories of the two trapped particles (1 pixel $=0.181 \mu$m) projected onto the $x$-axis (chosen along the line connecting the two traps) for different frames. Inset: a zoom of the particle position deviations inside the stationary trap.}
\label{Fig1}
\end{figure}

Consider two trapped particles. One (at position $\textbf{r}$) is held in a stationary trap (at the origin) and the other particle (situated at distance $\textbf{R}$ from the first particle) is brought closer in small steps (see Fig.\ref{Fig1}). The particle in the stationary trap will feel both the force $\mathbf F_{\mathrm {well}} (\textbf{r})$ exerted by its optical trap (due to a well potential $U_{\mathrm {well}}(\textbf{r})$) and a force ${\mathbf F}(R)$ due to the presence of the other particle. The additional force ${\mathbf F}$ causes the equilibrium position of the particle in the stationary trap to shift by a small amount. As the mobile trap is moved closer to the stationary one, the strength of this force increases and the particle deviates further. The particle in the mobile trap is similarly pushed in the opposite direction. For sufficiently large forces, Brownian motion can cause one of the particles to escape its trap. Figure \ref{Fig1} shows the positions of two particles as the mobile trap approaches, until one of the particles escapes. At large distances, the pair interaction between the particles is negligible, but as the traps approach, each particle deviates from its well center due to the electrostatic repulsion. If the trap potential is known, the equilibrium position of the particle in the trap can be calculated from the Boltzmann distribution. If we assume that the force $\textbf{F}$ exerted by the second particle does not vary significantly over the volume where the particle fluctuates; ${\mathbf F}(R) \approx {\mathbf F}(R\pm dr)$, where $dr$ is the typical fluctuation displacement of the particle inside the trap, the expected position is given by:
\begin{eqnarray}
\left\langle \textbf{r} \right\rangle = 
\frac{\int \textbf{r} ~ \exp \left( -\beta \left\lbrace  U_{\mathrm {well}}(\textbf{r}) - \textbf{F} \cdot \textbf{r} \right\rbrace \right) d \textbf{r}} {\int \exp \left( -\beta \left\lbrace  U_{\mathrm {well}}(\textbf{r}) - \textbf{F} \cdot \textbf{r}\right\rbrace  \right) d \textbf{r}},
\label{position-force}
\end{eqnarray}
where the integrations are carried out over the volume of the trap. After (numerically) calculating the integral, this relation can be inverted to calculate the force on a particle from its measured position. For perfectly harmonic traps, $ \textbf{F} \propto \left\langle \textbf{r} \right\rangle$.

To calculate the well potential we used the Mie-Debye representation given in Ref.\cite{mazolli,viana}. In these calculations, the parameters describing the laser trap configuration are the beam opening angle $\theta$ and $\gamma$ the ratio of the objective focal length to the beam waist $\omega_0$. We used $\theta = 64.245^\circ$, $\gamma =1.21$ to model our laser spot with the $100\times$ 1.4 NA lens with appropriate overfilling.\cite{horst2008, oostrum2009} The spherical aberrations, due to the refractive index mismatch between the coverslip and the  CHC, depend on the distance from the geometric focus to the glass surface (14 $\mu$m). To calculate the potential energy of the particle in the trap, the trapping force was integrated in the lateral direction at the trapping depth. The calculations show that for the parameters used in our experiments a distinct deviation from harmonic behavior, a relative increase in the stiffness at larger deviations, is expected.\cite{neves,richardson} For the measured forces with the largest deviations, this effect can be up to 20\%. By using the trapping potential from calculations we rest reassured that we do not underestimate these relatively larger forces. The trapping forces and potential assumed to scale with the power of the trap, such that the well potential is $U_{\mathrm {well}}(\textbf{r}) =\alpha U_{\mathrm {calc}}(\textbf{r})$. The prefactor $\alpha$ was measured from the standard deviation of the particle positions at large distances. 

\section{Results and discussion}
% *********************

\begin{figure}
\includegraphics[width=0.6\linewidth]{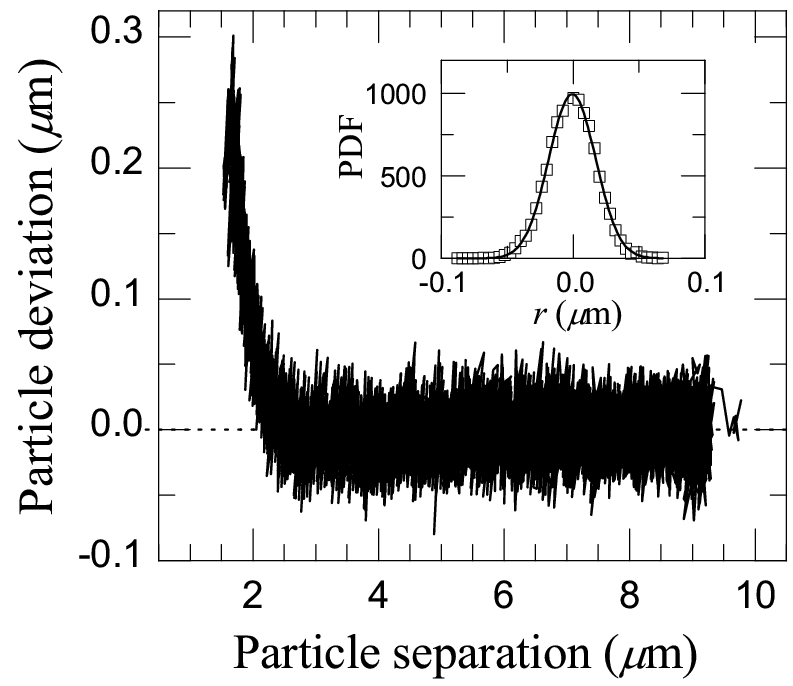}
\caption{Measured particle deviations inside the stationary trap plotted against particle separations for the sample with the highest salt concentration studied in this work. The particle starts feeling the approch of the mobile trap at a particle separation of about $2\sigma$. Inset: the probability distribution of displacements inside the stationary trap before any deviation and the fitted line is a Gaussian approximation to the experimental data.}
\label{Fig2}
\end{figure}

From our measurements, we plot the particle deviation in the stationary well $(r)$ versus the particle separation $(R)$ (for an example, see Fig.\ref{Fig2}). Initially, the particle exhibits Brownian motion inside the trap of a magnitude less than 0.1 $\mu$m, i.e. $<12\%$ of the particle diameter as demonstrated by the probability distribution of displacements plotted in the inset of the figure. From the standard deviation of this distribution we determine the proportionality factor $\alpha$, which sets the well depth. We then calculate average deviations at different interparticle distances by binning the data shown in Fig.\ref{Fig2}, and convert these to forces using Eq. (\ref{position-force}). Performing measurements with different laser powers (200 to 800 mW) gave the same results within our experimental resolution, which clearly demonstrates that possible light-induced particle interactions can be neglected.\cite{Polin,Burns-PRL89,Burns-Science90}

Our data of the pair-interaction forces are compared with the prediction of the linearized Derjaguin approximation\cite{Russel} with constant surface potential of the form 
\begin{equation}
F(R) =  \dfrac{k_{\mathrm B} T}{4\lambda_{\mathrm B}} ~\phi^{2} \kappa \sigma ~\dfrac{\exp(-\kappa \sigma (R/\sigma-1))}{1+\exp(-\kappa \sigma (R/\sigma-1))},
\label{YukawaCP}
\end{equation}
where $\phi=e\Psi_0/k_{\mathrm B} T$ is the dimensionless surface potential, $\lambda_B$ the Bjerrum length in CHC ($\simeq 7.3$ nm), $\kappa=\sqrt{4\pi \lambda_{\mathrm B} C_{\mathrm s}}$ the inverse Debye length assuming only monovalent ions of concentration $C_ {\mathrm s}$, $\sigma$ the particle diameter and $R$ the distance between the trapped particles. 

\begin{figure}
\includegraphics[width=0.8\linewidth]{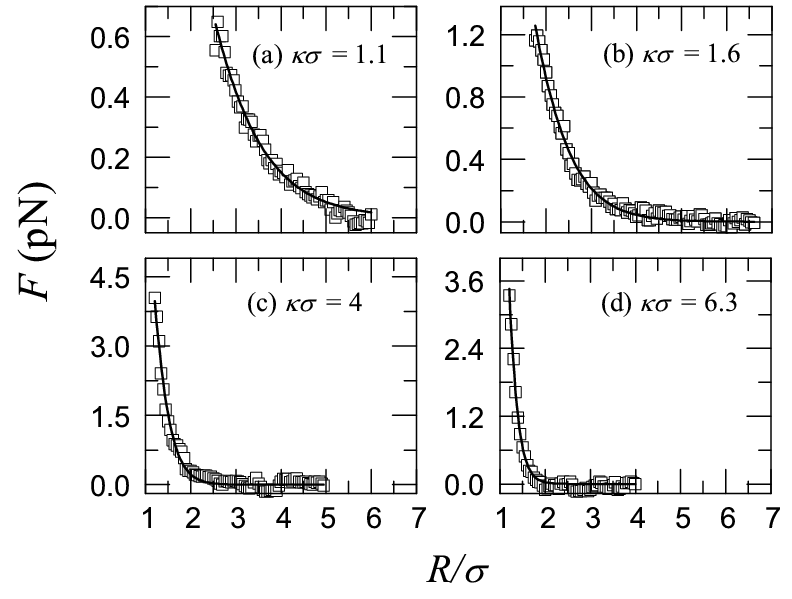} 
\caption{Interaction forces between charged colloidal PMMA particles suspended in cyclohexyl chloride with different salt concentrations resulting from different UV exposure times: ({\bf a}) initial purified sample. ({\bf b}) 1 hour UV exposure. ({\bf c}) 12 hours and ({\bf d}) 16 hours. The lines are fits based on the DLVO theory (Eq. (\ref{YukawaCP})) with fixed surface potential $\Psi_0 = 135$ mV, giving values for Debye screening lengths $\kappa^{-1}$ of 1.3, 0.87, 0.35 and 0.22 $\mu$m respectively.}
\label{Fig3}
\end{figure} 

Figure \ref{Fig3} shows the results for the different sets of measurements as the concentration of salt is increased by exposing the initial purified sample under an UV lamp at different exposure times: 0, 1, 12, and 16 hours respectively. When increasing the salt concentration, the range of the interparticle forces decreased. The data are well fitted with the DLVO force given by Eq. (\ref{YukawaCP}) with a fixed surface potential $\phi \simeq 5.3$ (or $\Psi_0 \simeq$ 135 mV) and a Debye screening length decreasing from $\kappa^{-1} \simeq 1.3$ $\mu$m in a purified sample to $\kappa^{-1}<0.2$ $\mu$m in the sample with the longest UV exposure time (highest salt concentration). The interaction force $F(R)$, initially soft and extremely long-ranged, becomes shorter-ranged for higher salt concentration, approaching that of hard-sphere systems. When performing the same measurements in the same sample with a different speed of approach, the resulting interactions do not differ significantly. This suggests that hydrodynamic effects due to the movement of the traps are negligible.

Since we do not follow the deviations in the $z$-direction but measure the projection of the particle separation onto the $xy$ plane, we slightly underestimate the distance between the two particles. In addition, the assumption that the force is approximatly constant over the integration volume in Eq. (\ref{position-force}) will lead to a small systematic errors in the measured forces. We performed Monte Carlo simulations to estimate these effects, simulating two particles in two optical traps with the interactions and well shape based on our experimental parameters. After analyzing the resulting data in the same way as the experimental data, the deviations in the obtained values for $\kappa \sigma$ and $\Psi_0$ were at most 3\% and 1\%, respectively.

\begin{figure}
\includegraphics[width=0.55\linewidth]{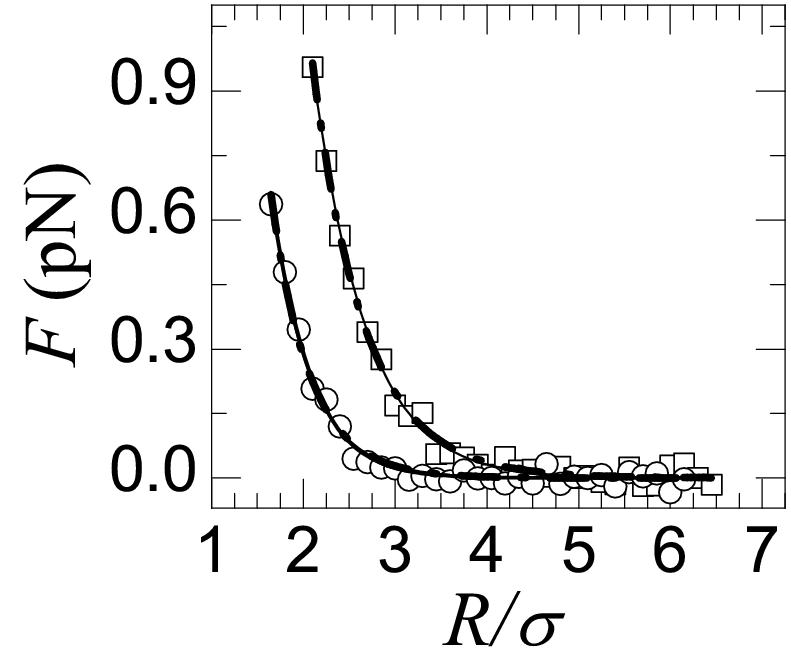} 
\caption{Interaction forces between charged colloidal PMMA particles in samples I and II with 0.026 and 0.26 $\mu$M added TBAC salt respectively. The lines are fits based on the DLVO theory with constant surface potential (full lines) with $\phi = 5.64$ and 3.37, giving values for screening lengths $\kappa \sigma$ of 1.86 and 2.58 respectively, and with constant surface charge (dashed lines) with $Q = 4.56$ and 2.55, giving values for screening lengths $\kappa \sigma$ of 1.6 and 2.06 respectively.}
\label{ForceWithTBAC}
\end{figure} 

\begin{figure}
\includegraphics[width=0.6\linewidth]{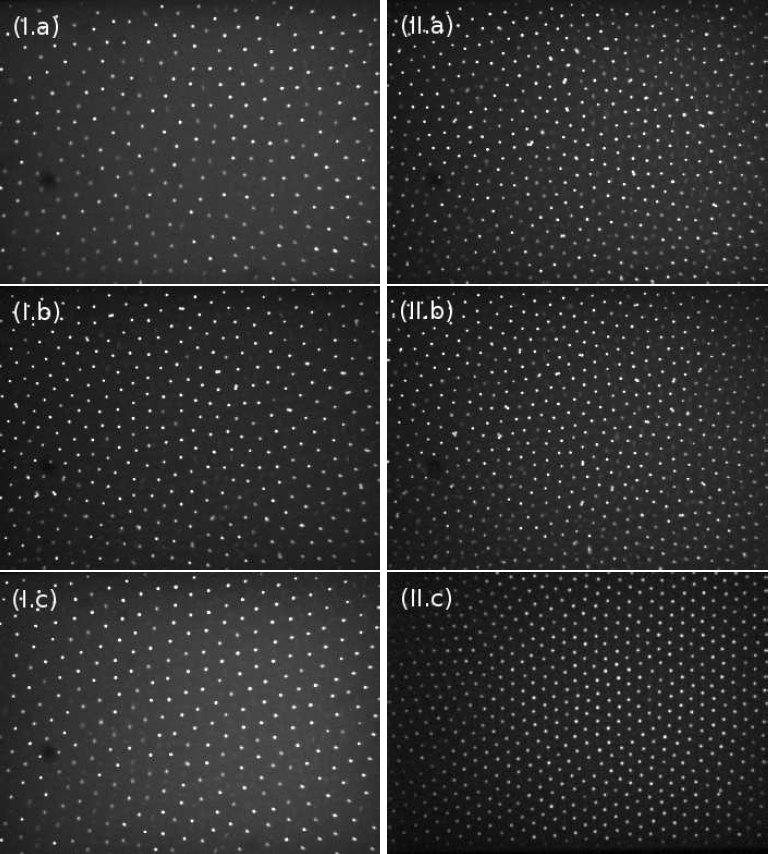}
\caption{Confocal images of the two sedimenting samples I and II with 0.026 and 0.26 $\mu$M added TBAC salt,  respectively, below (a), near (b) and above (c) freezing. The average particle-particle separations are 5.3 and 4.1 $\mu$m in I and II respectively. Using the coupling parameter needed for crystallization (Eq.\ref{CouplingParameterXtal}), this gives an estimate for the screening lengths $\kappa \sigma$ of about 2.3 and 2.8 respectively.}
\label{XtalSamplesIandII}
\end{figure} 

\begin{table}[t]
\centering
\begin{tabular}{|c|c|c|c|c|c|}
\hline
\multicolumn{6}{|c|}{Constant Surface Potential (Eq. (\ref{YukawaCP}))} \\ 
\hline
Sample &~~~ $\kappa \sigma$~~~ &~~~ $\phi$~~~ & ~~~$Z^*$~~~ &  ~~~$\phi_{\mathrm {el}}$~~~ & ~~~$Z^*_{\mathrm {el}}$~~~\\
\hline
I    &    1.86         &    5.64  & 1833  & 4.5       & 1130\\
II   &    2.58         &    3.37  & 910   & 2.42       & 585 \\
\hline
\multicolumn{6}{|c|}{Constant Surface Charge (Ref.\cite{Russel,YukawaCC})} \\ 
\hline
 &~~~ $\kappa \sigma$~~~ &~~~ $Q$~~~ & ~~~$Z^*$~~~ &  ~~~$\phi_{\mathrm {el}}$~~~ & ~~~$Z^*_{\mathrm {el}}$~~~\\
\hline
I    &    1.6         &    4.56  & 547   & 4.38       & 980 \\
II   &    2.06         &    2.55  & 237   & 2.43       & 510 \\
\hline
\end{tabular}
\caption{Measured parameters for the two samples I and II with 0.026 and 0.26 $\mu$M added TBAC respectively: the screening length $\kappa \sigma$, the dimensionless surface potential $\phi = e \Psi_0 / k_{\mathrm B} T$, the dimensionless surface charge $Q=eZ^{*}/4\pi \varepsilon k_{\mathrm B} T \kappa^{-1}$, and the effective charge $Z^{*}$. The effective charges quoted for the constant potential fits are for infinite particle separation. The last two columns are based on the electrophoresis measurements.}
\label{valuesPMMACHCTBAC}
\end{table}

To further test the reliability of this method, we conducted two more measurements using two other dispersions containing a controlled amount of salt tetrabutylammonium chloride (TBAC, Sigma-Aldrich). We independently estimated the Debye screening lengths from the scaling law for the liquid-solid phase transition in Yukawa systems,\cite{YukawaPhaseDiagram} and measured the particle surface charges by means of electrophoresis. We prepared a solution of TBAC in purified CHC, which we allowed to equilibrate for a week. We then filtered the saturated solvent ($\sim$260 $\mu$M TBAC) and diluted it by adding an amount of purified CHC to prepare two solvents I and II containing 0.026 and 0.26 $\mu$M of TBAC, respectively. The measured force-distance profiles of PMMA particles in these two salt added solvents are again well fitted with the DLVO force but with different surface potentials. The data are also well fitted with the constant surface charge formula\cite{YukawaCC, Russel} (see Fig.\ref{ForceWithTBAC} and Table \ref{valuesPMMACHCTBAC} for the different parameters). When the two samples are left vertically, sedimentation of particles induced crystallization. The samples were imaged with a tilted confocal microscope that allows to scan at all heights. The average particle-particle separation $d$ for the two systems near freezing were 5.3 and 4.1 $\mu$m respectively (see Fig.\ref{XtalSamplesIandII}). Using the scaling law proposed in Ref.\cite{YukawaPhaseDiagram}, the coupling parameter needed for crystallization: 

\begin{equation}
\frac{U(d)}{k_{\mathrm B} T} \left( 1+\kappa d + \frac{\left( \kappa d \right)^{2}}{2}\right) =106.6, 
\label{CouplingParameterXtal}
\end{equation}
where $U(d)$ is the potential energy at the typical particle-particle separation. Using surface potentials from our force fits, this gives values for $\kappa \sigma$ of about 2.3 and 2.8 respectively, in the same order of magnitude as from our force measurements. In addition, we obtained independent measurements of the ionic strength from the conductivity of the particle-free solvents. A commercial conductivity meter (Scientifica 627) was used to measure the conductivity of the two solvents with the added salt. This yielded conductivities of 1600 and 9800 $p$S/cm respectively. Using Walden's rule,\cite{Walden} the corresponding screening lenghts are $\kappa \sigma=2.1$ and $5.5$ respectively. The screening parameters obtained from conductivity measurements
are on the same order as those obtained from our force measurements and
Eq.\ref{CouplingParameterXtal}, although the value for sample II is slightly higher.
By fixing $\kappa\sigma = 5.5$ for sample II, we were neither able to
fit our force measurement data assuming a constant surface
potential nor a constant surface charge. Therefore, we believe
that the estimated value $\kappa\sigma = 2.8$ from Eq.\ref{CouplingParameterXtal} is more accurate in this case.

Electrophoretic measurements on the same dilute samples were conducted by driving the particles in a dc-electric field ($E \approx 1-3$ V/mm) and measuring the electrophoretic mobility from particle tracking of the confocal images. Using the values of the Debye screening length $\kappa^{-1}$ from our force fits and the measured mobilities, the surface potentials were obtained using recent calculations for electrophoresis.\cite{Carrique} This gives values of $\phi_{\mathrm {el}}=4.5$ and 2.42 for samples I and II respectively when using the screening lengths from the constant surface potential fits and, 4.38 and 2.43 for the case of constant surface charge. Finally, to translate the surface potential into a particle charge $Z^{*}$, we used the empirical relationship proposed by Loeb \textit{et al.}\cite{Loeb} The results of the electrophoresis measurements are slightly lower than the measured force parameters with a constant surface potential but slightly higher than the parameters obtained under the assumption of a constant surface charge. These results suggest the importance of charge regulation in the case of PMMA dispersed in CHC with added TBAC.\cite{RoyallJCP2006}

\section{Conclusion}
% *********************
In summary, we have measured the interaction forces acting between charged PMMA colloidal particles suspended in a relatively low dielectric medium directly from the deviations of particle positions inside an optical trap. The measured forces include all possible effects such as the influence of the surrounding particles. Confocal microscopy can be used to track the particles at relatively high volume fractions $\sim$ 15 $\%$ where optical tweezers can still be operated. As a result, this method allows to measure particle interactions at higher volume fractions. Additionally, our measurements quantified for the first time the effect of exposing a PMMA/CHC dispersion to UV light. When increasing the salt concentration by exposing the sample to UV light, the force measurements agree well with the classical DLVO theory assuming a constant surface potential. On the other hand, when adding tetrabutylammonium chloride (TBAC) to vary the salt concentration, surface charge regulation seemed to play an important role. Future work will explore the effect of particle concentration and the nature of the interactions in dense systems.

\section*{Acknowledgements}
% *********************
We are grateful to Ren\'e van Roij for helpful discussions, Johan Stiefelhagen for particle synthesis and building the electrophoretic cells, Peter Helfferich for technical support and Peter Lu for the automated acquisition software. This work was supported by NWO-SRON. FS and MD acknowledge the support of an NWO-VICI grant. TV acknowledges the support of an NWO-CW grant.

% *********************


\begin{thebibliography}{999}
\bibitem{YukawaCC}
The linearized Derjaguin approximation of the pair-interaction forces at constant surface charge is given by 
$F(R) =  \dfrac{k_B T}{4\lambda_B}~Q^2~\kappa \sigma ~ \dfrac{\exp(-\kappa \sigma (R/\sigma-1))}{1-\exp(-\kappa \sigma (R/\sigma-1))}$, where $Q=eZ^{*}/4\pi \varepsilon k_{\mathrm B} T \kappa^{-1}$ is the dimensionless surface charge.
\end{thebibliography}
\end{document}